# Effect of Nanoparticle Size on the Morphology of Adsorbed Surfactant Layers

*Dersy Lugo,[†] Julian Oberdisse,[‡] Alain Lapp[§] and Gerhard H. Findenegg*[†]*

[†]*Stranski Laboratorium für Physikalische und Theoretische Chemie, Technische Universität Berlin, Strasse des 17. Juni 124, D-10623 Berlin, Germany.*

[‡]*Laboratoire des Colloïdes, Verres et Nanomatériaux, UMR 5587 CNRS, Université Montpellier II, 34095 Montpellier, France*

[§]*Laboratoire Léon Brillouin, CEA Saclay, 91191 Gif-sur-Yvette CEDEX, France*

*Corresponding author: findenegg@chem.tu-berlin.de

**Abstract**

The surface aggregates structure of dimethyldodecylamine-*N*-oxide ($C_{12}$DAO) in three silica dispersions of different particle sizes (16 - 42 nm) was studied by small-angle neutron scattering (SANS) in a $H_2O/D_2O$ solvent mixture matching the silica. At the experimental conditions (pH 9) the surfactant exists in its nonionic form and the structure of the adsorbed layer is not affected by added electrolyte. It is found that $C_{12}$DAO forms spherical surface micelles of 2 nm diameter on the 16 nm silica particles, but oblate ellipsoidal surface micelles are formed on the 27 and 42 nm particles. The dimensions of these oblate surface aggregates (minor and major semi-axes $R_n$ and $R_{lat}$) are similar to those of $C_{12}$DAO micelles in the aqueous solutions. It is concluded that the morphological transition from spherical to ellipsoidal surface aggregates is induced by the surface curvature of the silica particles. A comparison of the shape and dimensions of the surface aggregates formed by $C_{12}$DAO and $C_{12}E_5$ on the 16 nm silica particles demonstrates that the nature of the surfactant head group does not determine the morphology of the surface aggregates, but has a strong influence on the number of surface aggregates per particle, due to the different interactions of the head groups with the silica surface.

# 1. Introduction

Surfactants play an important role in many industrial processes involving colloidal dispersions, as their adsorption onto the particles often leads to enhanced colloid stability. A structural characterization of this adsorbed layer is a prerequisite for gaining a better understanding of its mode of operation in stabilizing or flocculating dispersion. Adsorption isotherms of nonionic surfactants on hydrophilic (oxide) surfaces commonly exhibit a pronounced sigmoidal shape, i.e., a low-affinity initial region followed by a region in which the adsorption increases steeply and reaches a plateau near the critical micelle concentration ($CMC$).[1] This behavior suggests that adsorption represents a surface aggregation similar to micelle formation in solution. Scanning probe microscopy (AFM) studies at planar surfaces indicated that either laterally uniform surfactant bilayers or small surface micelles may be formed, depending on the nature of the surfactant head group and the degree of hydrophilicity of the solid surface.[2] The nature of the surfactant layers adsorbed on colloidal particles in aqueous dispersions was studied by small-angle neutron scattering (SANS), which allows to highlight the adsorbed layer against a uniform scattering background by matching the colloidal particles with a partially deuterated aqueous solvent.[3,4,5,6,7] In the earlier of these studies the adsorbed surfactant was modeled as a laterally uniform layer,[3,4] but the existence of discrete micellar aggregates at the surface of the silica particles ('micelle-decorated silica beads') has been reported more recently.[5,6,7]

Recently we reported that the surfactant penta(ethyleneglycol) monododecyl ether ($C_{12}E_5$) is adsorbed in the form of individual spherical surface aggregates on silica nanoparticles of 16 nm diameter,[7] in agreement with earlier findings for the surfactant Triton X-100 on Bindzil-type silica particles of similar size.[5,6] This finding is remarkable in view of the fact that $C_{12}E_5$ prefers aggregates of lower mean curvature, viz., worm-like micelles in aqueous solutions,[8,9] and a laterally homogeneous bilayer on planar hydrophilic silica surfaces.[2a] We conjectured that the preference for small surface micelles is a consequence of the high surface curvature of the silica nanoparticles, which prevents an effective packing of the hydrophobic tails in an adsorbed bilayer, whereas a favourable packing of the tails is possible in a spherical micelle. In order to test this concept and to find out to what extent the structure of the adsorbed layer at the surface of the silica nanoparticles depends on the size and chemical nature of the surfactant head group it was of interest to extend this study to a different class of nonionic surfactants. On the other hand, it was of interest to study the influence of size of the silica nanoparticles on the surface aggregate structure of the surfactant.

The present study was performed with dimethyldodecylamine-*N*-oxide ($C_{12}DAO$), an amphoteric surfactant that exists in a zwitterionic (net non-ionic) form at pH above 7, but in a cationic form at low pH due to a protonation of the head group. $C_{12}DAO$ has a much smaller head group of less hydrophilic character than $C_{12}E_5$.[10] Phase diagrams, thermodynamics and self-assembly structures of aqueous systems of alkyl dimethylamine oxides have been extensively studied,[11] and the interaction of alkyl DAO systems with hydrophilic and hydrophobic surfaces was investigated by adsorption calorimetry[12,13a] and streaming potential measurements.[13b] Based on the adsorption enthalpy results, Pettersson and Rosenholm[13a] concluded that the adsorption mechanism at the solution/silica interface of $C_{12}DAO$ in its nonionic form is different from that in the protonated form, and they speculated that in the nonionic form $C_{12}DAO$ forms ellipsoidal aggregates; while in the protonated form $C_{10}DAO$ and $C_{12}DAO$ are likely to form spherical surface micelles. The conclusion about the formation of spherical surface micelles by $C_{10}DAO$ on silica was consistent with the sorption enthalpy results of Király and Findenegg.[12] However, in neither of these studies direct information about the surface aggregate structures was obtained. In the present work we use SANS to clarify the structure of the adsorbed layer of $C_{12}DAO$ on silica nanoparticles of three different sizes (16 to 42 nm diameter), with a focus on the effect of particle size on the type of surface aggregate formed.

## 2. Experimental Section

### 2.1 Materials

*N,N*-Dimethyldodecylamine-*N*-oxide, $C_{12}DAO$ (Fluka, purity ≥ 98%), tetraethyl orthosilicate, TEOS (Fluka, purity ≥ 99.0%), ammonia (Sigma-Aldrich, A.C.S. reagent, 30-33% in water), Ethanol, $C_2H_5OH$ (Berkel AHK, purity ≥ 99.9%), and $D_2O$ (Euriso-top, 99.9% isotope purity), were used without further purification. Reagent-grade water was produced by a Milli-Q 50 filtration system (Millipore, Billerica, USA) and additionally passed through a 0.22 μm membrane to remove micrometer-sized particles. The colloidal silica suspensions, Ludox SM-30 (30 wt-% in water) and Ludox HS-40 (40 wt-% in water) were supplied by Sigma-Aldrich. They were dialyzed with reagent-grade water (2 weeks) and filtered with a 0.8 μm Millipore Steril Filter. The silica concentration in the purified suspensions was about one half of the original concentration. Their pH was adjusted to 9 to preserve colloidal stability.

**2.2 Sample Preparation**

**Preparation and Characterization of silica nanoparticles.** Three samples of monodisperse silica nanoparticles were prepared by two variants of the Stöber synthesis.[14] Silica I (diameter 16 nm) and silica II (27 nm) were made by particle growth from Ludox SM-30 and Ludox HS-40 dispersions, respectively, using the procedure described earlier.[7] Silica III (diameter 42 nm) was prepared by condensation of TEOS starting from a mixture of 100 mL ethanol and 7.5 mL ammonia at 60 °C, and 3.0 mL TEOS was added dropwise in a 250 mL three-neck round flask equipped with a magnetic stirrer and reflux-condenser. Particle growth was allowed to continue for 24 h at 60 °C. The excess of ethanol and ammonia was then removed from the resulting suspension in a rotary evaporator (40°C, 160 mbar) by reducing the volume to 20% of the initial value. The suspension was dialyzed with reagent-grade water for 1 week, filtered and stored at pH 9 in a refrigerator at 280 K.

**Sample preparation for SANS.** Dilute silica dispersions (0.4 to 1.5 vol.-%) in a contrast-matching H$_2$O/D$_2$O mixture of scattering length density $\rho = 3.54 \times 10^{10}$ cm$^{-2}$ were prepared for the SANS measurements, as determined experimentally in the earlier work.[7] The total surface area of silica in the samples was obtained from the mass fraction of silica in the dispersions (determined gravimetrically) and the specific surface area $a_s$ of the silica. Samples with different adsorbed amounts of C$_{12}$DAO, were prepared by adding appropriate amounts of the surfactant directly to the aqueous dispersion. The adsorption isotherm of C$_{12}$DAO on Davisil silica gel reported by Pettersson and Rosenholm[13a] was used to estimate the amounts of surfactant needed for a given surface concentration on the silica particles. According to that work, a plateau value $\Gamma_{mx} = 7.5$ µmol m$^{-2}$ is reached at a solution concentrations somewhat above the *CMC* (~ 2 10$^{-3}$ M), and the surface concentration at the *CMC* is about ⅞$\Gamma_{mx}$. This value was chosen as the highest surface concentration of the surfactant on the silica particles, to avoid free micelles in the solution. All samples were kept at pH 9 to minimize particle aggregation.

**2.3 Methods**

**Small-angle neutron scattering.** Experiments were carried out on the spectrometer PAXY (Laboratoire Léon Brillouin, Saclay, France). Scattering profiles were taken in a range of the scattering vector $q$ from 0.03 to 3 nm$^{-1}$ using three wave-length/sample-to-detector distance configurations: wave length $\lambda = 6$ Å with sample-to-detector distances 1 m and 5 m (collimation distances 2.5 and 5 m); and $\lambda = 12$ Å, sample-to-detector 5 m (collimation distance 5 m). Samples were contained in standard 2-mm path length quartz cells (QS, Hellma), thermostated at 25.3 ± 0.1 K. Intensities were divided by transmission and sample thickness (1 and 2 mm), empty cell scattering was subtracted and detector calibration achieved by dividing by 1 mm $H_2O$ scattering. Absolute units ($cm^{-1}$) were obtained by measuring the incident flux and using a standard procedure.[15] The incoherent scattering background of the samples was subtracted by enforcing a high-$q$ Porod ($I(q) = Pq^{-4}$) behaviour, with background intensity values typical for $H_2O/D_2O$ mixtures.

**Transmission Electron Microscopy.** TEM images were taken using a Tecnai $G^2$ 20 S-Twin electronic microscope operating at an accelerating voltage of 200 kV and an electron source of $LaB_6$. Samples for TEM were first diluted to 0.8 wt-% in ethanol and then prepared by drying a droplet of the dilute dispersion on a copper grid (coated with a carbon film with a thickness of 20 nm). TEM images were taken at a minimum of 5 different locations on the grid, and a total of 220 particles were measured per sample to ensure good statistics in the determination of the particles size.

**Nitrogen Adsorption.** The specific surface area of the silica samples was determined by nitrogen adsorption at 77 K using a Gemini III 2375 Volumetric Surface Analyzer (Micromeritics). For this purpose the silica dispersion was dried at 218 K for two days using a Freeze Dryer Alpha 2–4 LD/plus (Martin Christ), and then outgassed at 393 K for 1 h under vacuum.

**Zeta Potential.** Zeta potential measurements were carried out with a Malvern Zeta-Sizer 2000 using the diluted silica particle dispersions at pH 9 and at 298 K.

## 3. Results and Discussion

### 3.1 Characterization of the silica sols

The silica samples were characterized by transmission electron microscopy (TEM), small-angle neutron scattering (SANS), zeta potential and nitrogen adsorption measurements. TEM images indicate that silica I and II, which were obtained by silica deposition on Ludox, have a wider size distribution than silica III, which was prepared by direct Stöber synthesis (Figure 1). The average particle radius, $R_{TEM}$, and its standard deviation, $SD_{TEM}$, were determined by Gaussian fits to the histograms in Fig. 1 (see Table 1). The silica particles prepared by overgrowth of Ludox (silica I and II) have a higher zeta potential than silica III at pH 9 (see

Table 1), indicating a somewhat different surface-chemical behavior of the two types of silica. The zeta potentials suggest that all silica dispersions are electrostatically stabilized.

Quantitative information about the particle size and size distribution of the silica sols was obtained by SANS. The scattering profile of a dilute (1.5 vol.-%) dispersion of silica II in $D_2O$-rich water is shown in Figure 2. In this case $I(q)$ reflects the form factor of the particles. It is characterized by a Guinier regime at low values of the scattering vector $q$, a characteristic oscillation at $q \approx 0.4$ nm$^{-1}$ which relates to the radius of the silica particles, and a Porod-law behaviour, $I(q) \sim P \cdot q^{-4}$ at high $q$. The entire scattering profile can be represented by the form factor of spheres with log-normal size distribution, characterized by a radius $R_S$ and polydispersity $\sigma$.[16] Values of $R_S$ and $\sigma$, average radius $<R_S>$, average surface area $<A_S>$ and average volume $<V_S>$ of the particles are given in Table 2. The scattering profiles of all three silica sols, fits with the log-normal size distribution of spheres, and details of the data analysis are given in the Supporting Information (SI).

The specific surface area $a_s$ of the silicas was determined from the nitrogen adsorption isotherms by the BET method.[17] Linear BET plots (correlation coefficient ≥ 0.9996) were found for relative pressures $p/p_0$ from 0.05 to 0.3 for the three samples. (Adsorption isotherms and BET plots are shown in SI). Values of $a_s$ and the geometric surface area $a_{geom} = 3/\rho_S R_S$ derived from the particle radius $R_S$ (from SANS) and the mass density of silica $\rho_S$ (2.20 g cm$^{-3}$) are given in Table 3. Values of $a_S/a_{geom}$ between 1 and 2 are obtained, increasing with the particle radius. This trend indicates that the surface roughness of the particles increases with size. Silica I ($a_S/a_{geom}$ = 1.14) has a very low surface roughness, while the value $a_S/a_{geom}$ = 1.85 for silica III indicates significant roughness (surface corrugations at a periodicity $l \approx 1$ nm and profile depth $d \approx l$). Alternatively, the result for silica III may be explained by a moderate degree of microporosity.

### 3.2 SANS study of adsorbed $C_{12}DAO$ layer

Scattering profiles $I(q)$ for $C_{12}DAO$ in the absence and presence of silica particles are shown in Figure 3. These SANS measurements (and those presented in the later figures) were made in contrast-matching $H_2O/D_2O$, so that the scattering contrast is solely due to the surfactant. Accordingly, the difference in the scattering profiles obtained in the absence and presence of silica in Fig. 3 must be due to a different organization of the surfactant. As discussed later, the scattering profile in the absence of silica is indicative of surfactant micelles of ellipsoidal shape, while the scattering profile in the presence of silica indicates that the surfactant forms an adsorbed layer on the silica particles. Figure 3 also shows that addition of an electrolyte (0.1 M NaBr) to the silica-containing system is causing no significant changes of the scattering profile, indicating that the electrolyte does not affect the adsorption at the silica surface. The profiles in Fig. 3 were obtained with silica II at a surfactant concentration corresponding to a nearly complete adsorbed layer (⅞$\Gamma_{mx}$), but analogous results were found with silica I in the absence and presence of 0.1 M NaBr. The absence of a salt effect on the adsorption is expected because at the given pH $C_{12}$DAO behaves as a nonionic surfactant with negligible degree of protonation.[13]

Scattering profiles for different surface concentrations of the surfactant (¼$\Gamma_{mx}$, ½$\Gamma_{mx}$, ¾$\Gamma_{mx}$ and ⅞$\Gamma_{mx}$) on silica I and II, and for ⅞$\Gamma_{mx}$ on silica III are presented in SI. Qualitatively, all scattering profiles are similar to those for $C_{12}$DAO on silica II shown in Fig. 3, but significant differences in detail can be found, as will be shown below. The local maximum in $I(q)$ appears at $q_{max} \approx 0.42$ nm$^{-1}$ (silica I), 0.30 nm$^{-1}$ (silica II), and 0.20 nm$^{-1}$ (silica III). The overall scattering intensity as well as the relative height of the maximum at $q_{max}$ both increase with the surface concentration of $C_{12}$DAO. The analysis of the SANS profiles was made in two steps: Initially, simple geometrical modelling was used to estimate the volume, effective layer thickness and volume-based surface area of the adsorbed $C_{12}$DAO. In the second step, nonlinear least-squares fitting of the scattering data to appropriate structure factor models was employed in order to extract information about the size and shape of the surface aggregates.

**3.3 Geometric Modelling.**

A model-free analysis of the Guinier and Porod regimes of $I(q)$ in terms of the dry volume, layer thickness, and volume-based surface area of the adsorbed surfactant was performed as a basis for simple geometrical models of the surface aggregates.

The Guinier expression $I(q) = I_0 \exp(-R_g^2 q^2/3)$ can be used to fit the data in the low-$q$ region. For example, for silica II ($R_G = 15.8$ nm) we find a radius of gyration $R_g = 17.3$ nm at the highest surface concentration (⅞$\Gamma_{mx}$). In the contrast-match scenario of our experiment, $R_g$ must have a value between the silica radius $R_G$ and $R_G + \delta$, depending on the surfactant density profile. Assuming for simplicity that $R_g$ is half-way between these two values, a typical layer thickness of the adsorbed surfactant is $\delta = 2 \cdot (17.3 - 15.8) = 3$ nm. From the high-$q$ region (Porod regime) we obtain the volume-based surface area $S/V$ of the surface aggregates, since the concentration of free micelles in solution is negligible at the chosen surfactant concentrations. The respective value for free micelles can be derived from the scattering profile of the surfactant in the absence of silica (Fig. 3). One finds that $S/V$ for the free micelles is about 10% lower than for the surface aggregates. The similar magnitude of the two values implies similar morphologies of the surfactant aggregates in solution and on the surface. This excludes adsorbed half-micelles,[13] which would require considerably more surface area.

For non-interacting particles, the dry volume of adsorbed $C_{12}DAO$ per silica bead, $V_{dry}$, can be derived from the scattering cross section at zero angle by the relation $I_0 = \varphi \Delta\rho^2 V_{dry}$, where $\varphi = 0.00891$ is the volume fraction of $C_{12}DAO$ in the dispersion and $\Delta\rho = 3.72 \times 10^{-4}$ nm$^{-2}$ is the scattering contrast between surfactant and background. From $V_{dry}$ and the mean particle radius $<R_S>$ (Table 2) one can determine the effective layer thickness $\delta$ of dry surfactant. Values of $V_{dry}$ and $\delta$ derived from the SANS data in this way are given in Table 4. They are compared with values estimated from the adsorption isotherm of Pettersson,[13] using the mass densities 0.88 g cm$^{-3}$ (pure $C_{12}DAO$) and 2.20 g cm$^{-3}$ (silica), and the values of the mean surface area per silica particles $<A_S>$ (Table 2). Reasonable agreement between the two sets of values is found for most samples, but large deviations appear for the sample with silica III. The results in Table 4 indicate that at surface concentrations up to ½$\Gamma_{mx}$, the effective layer thickness $\delta$ is significantly smaller than the extended tail length of $C_{12}DAO$ ($l_c = 1.67$ nm)[18] while at surface concentrations above ½$\Gamma_{mx}$, the layer thickness approaches $l_c$. This suggests the existence of either a monolayer (which is physically implausible at hydrophilic surfaces) or patches of bilayer.[19] Simple geometric modeling based on surface area and volume of adsorbed surfactant (from the Porod constant and $I_0$, respectively) indicates that these patches must have dimensions close to micelles. The possibility of discrete surface micelles as reported recently[5,6,7] thus appears plausible. On the assumption that the volume of such surface micelles is similar to that of micelles in solution, the number of surface micelles $N_{mic}$ can be estimated by dividing the dry volume of adsorbed surfactant by the volume of a free micelles. Values of $N_{mic}$ obtained in this way are given in Table 4.

### 3.4 Core-shell model

The spherical core-shell model[3,7,20] was adopted to see if the scattering profiles are consistent with a laterally homogeneous surfactant layer. Three different values of the layer thickness (1.6, 3.2, and 4.0 nm) were tested in the modeling of the data for high surface concentrations of $C_{12}DAO$: The first value corresponds to the effective thickness, $\delta_{eff}$, as obtained for high surface concentrations from the simple geometric analysis (Table 4); the second value is the expected bilayer thickness, *i.e.* twice the monolayer thickness, and the third value represents the mean thickness of a bilayer of $C_{12}E_5$ at the surface of silica I.[7] A fit of the data for the surface concentration ⅞$\Gamma_{mx}$ on silica II with the three values of layer thickness is shown in

Figure 4. In these fits the polydispersity in radius of the silica particles is taken into account as explained in our earlier work.[7]

The fits based on the expected layer thickness $\delta$ (3.2 nm) or a greater thickness (4.0 nm) exhibit similar features as noted in the preceding studies with the surfactants Triton X-100[5] and $C_{12}E_5$[7]: The model gives a good fit to the scattering data in the low-$q$ regime including the maximum at intermediate $q$, if the theoretical intensities are multiplied by a scale factor $f = 0.22$ ($\delta = 3.2$ nm) or $f = 0.12$ ($\delta = 4.0$ nm) (*cf.* inset Fig. 4). As the scattered intensity is proportional to the surfactant volume on each bead, this implies that the core-shell model with such a film thickness considerably overestimates the adsorbed surfactant volume. On the other hand, a fit with the layer thickness $\delta_{eff} = 1.6$ nm reproduces all features of $I(q)$ without any scale factor $f$, except for the shoulder at $q \approx 1.0$ nm$^{-1}$, causing some underestimate of the specific surface area of adsorbed surfactant. In addition, the maximum in $I(q)$ appears at somewhat too high $q$, indicating that the real value of the average layer thickness should be greater than 1.6 nm. The agreement with the experimental data for the film thickness 1.6 nm is of course related to the fact that in this case the volume of adsorbed surfactant is conserved. However, although the core-shell model with a layer thickness 1.6 nm gives a good representation of the scattered intensities, the result is unrealistic because this value of $\delta$ corresponds to only about half the thickness expected for a bilayer of $C_{12}DAO$ at the solid/solution interface. Results similar to those shown in Fig. 4 were also obtained for lower surface concentrations of $C_{12}DAO$ on silica II and for the adsorption of $C_{12}DAO$ on silica I.

The shortcomings of the core-shell model suggest that the adsorbed surfactant does not form a laterally uniform layer but smaller surface aggregates, such as spherical or ellipsoidal surface micelles, which have a higher surface area at a given total adsorbed volume. Accordingly, models of silica particles decorated with such small surface micelles were applied to the present data, as described below.

### 3.5 Micelle-decorated silica model

In previous publications[5,6,7] we have developed and applied a form factor model for objects made of small spherical micelles adsorbed on an indexed-matched silica bead. Parameters of the model are the radius of the silica bead $R_S$, the radius of spherical surface micelles $R_{mic}$, and number of surface micelles $N_{mic}$ per silica particle, as well as a polydispersity parameter. Polydispersity of the silica bead is accounted for by performing the calculation for silica radii drawn from a distribution function, and averaging. For a given silica bead of radius $R_S$, the micelle centers are supposed to sit on a spherical shell of radius $R_S + R_{mic}$. The excluded volume of the spherical micelles is determined by their radius $R_{mic}$, which also acts as a lateral interaction parameter for spheres. It is assumed that the number density of spheres in the layer is independent of bead radius and that $N_{mic}$ corresponds to the number of micelles on a bead of average radius. The algorithm consists of the following steps: (i) positioning the micelles in a random manner on the shell, possibly allowing for lateral reorganization following a Monte Carlo motion; (ii) calculation of the micelle-micelle structure factor using the Debye formula; (iii) calculation of the scattered intensity in absolute units; and (iv) convolution with the resolution function of the spectrometer.

In order to check if a model of *spherical* micelles is consistent with the data, their radius and number was estimated directly from the scattering curves. The micellar radius must be approximately half the thickness of the layer, and $N_{mic}$ and $R_{mic}$ are related by volume conservation to the amount of adsorbed surfactant as determined either by the adsorption isotherm or by the low-$q$ fit of the core-shell model with factor $f$. Similarly, the amount of surface produced by $N_{mic}$ spheres of radius $R_{mic}$ must match the volume-related surface area determined from Porod's law. This constrains the model considerably, and parameters can only be varied in a narrow range.

Fitting of the scattered intensities with this model and the estimation of the real adsorbed surfactant volume, $V_{tot}$, was based on the results of the core-shell model for the layer thicknesses $\delta$ = 3.2 and 4.0 nm. The real adsorbed surfactant volume was determined by introducing the effective volume fraction of surfactant in the shell, $X$ (i.e., fraction of the layer volume occupied by the surfactant), which is related to the scale factor $f$ by $X = \sqrt{f}$. For example, for silica II at the surface concentration $\frac{7}{8}\Gamma_{mx}$, the scale factor $f = 0.22$ introduced for the layer thickness 3.2 nm implies that only 47% of the layer volume is occupied by $C_{12}DAO$. The total surface area of the adsorbed surfactant, $A_{tot}$, was calculated from the volume-related surface area of adsorbed surfactant as determined by Porod's law, $(S/V)_{surf} = P/2\pi\Delta\rho^2$, and the number density of silica beads, $(N/V)_S = \varphi/\langle V_S \rangle$, by the relation $A_{tot} = (S/V)_{surf}/(N/V)_S$.

The number and dimensional parameters of surface aggregates of different morphologies were estimated from the real volume, $V_{tot}$, and total surface area, $A_{tot}$, of adsorbed surfactant. The dimensional parameters of surface aggregates of given shapes depend strongly on the layer thickness $\delta$ used to calculate $V_{tot}$. At first we assumed the formation of isolated spherical aggregates of radius $R_{mic}$ and number of micelles $N_{mic}$. In most cases a fits with a layer thickness $\delta$ = 4.0 nm (i.e., a value similar to that found by the Guinier approximation, Section 3.3) gave a somewhat better fit than with $\delta$ =3.2 nm, although the difference was within a 5% in most cases. Results for fixed $\delta = 4.0$ nm are summarized in Table 5. A noteworthy finding is that the number of surface aggregates estimated by this model ($N_{mic}$) is similar to that obtained by the simple geometrical analysis ($N_{mic}$, *cf*. Section 3.3). This indicates that the simple geometric analysis gives reliable information about the morphologies of the surface aggregates.

**C$_{12}$DAO on silica I.** Figure 5 shows the scattering data for C$_{12}$DAO on silica I at the surface concentrations ¾Γ$_{mx}$ and ⅞Γ$_{mx}$, and fits with the micelle-decorated silica model. In this case, a good fit of the scattering profile was obtained by assuming that C$_{12}$DAO is adsorbed in form of isolated spherical surface micelles of radius $R_{mic} = 2.0$ nm. The good representation of the data in the high-$q$ region supports the conjectured uniform size of the surface aggregates, and the fit of the data at intermediate $q$ indicates that the increasing amplitude of the maximum at $q_{max}$ can be explained by an increasing number of surface aggregates as the surface concentration is increased (*cf*. Table 5). Some deviations between the experimental and predicted $I(q)$ appear in the $q$ regime just below $q_{max}$, where the model somewhat overestimates the total volume of the adsorbed surfactant aggregates (see Fig. 5). The strong increase of $I(q)$ at the lower end of the experimental $q$ range is a hint of a maximum in $I(q)$ at lower $q$, which was not captured because measurements at smaller angles were not performed for this silica. Such a maximum indicates repulsive interaction between the silica beads coated with small surface aggregates of C$_{12}$DAO. The quality of the fit of the low-$q$ region could not be improved by decreasing $N_{mic}$, the number of surface micelles, at fixed radius $R_{mic} = 2.0$ nm (*cf*. Table 5), since this implies a decrease in the surfactant volume fraction in the layer and thus lowers the amplitude of the maximum at $q_{max}$, which is inconsistent with the experimental $I(q)$. Similarly, no better fit was found when $R_{mic}$ was increased at a fixed value of $N_{mic}$.

**C$_{12}$DAO on silica II and III.** The scattering data for C$_{12}$DAO on silica II and III were also analysed in terms of the model of spherical surface aggregates. However, for these silicas reasonable fits with spherical surface micelles could be obtained only by adopting unrealistic values of the micelle radius $R_{mic}$. Specifically, with silica II a surface micelle radius $R_{mic} = 0.85$ nm was obtained for low surface concentrations (¼Γ$_{mx}$, and ¼Γ$_{mx}$) and an even lower value ($R_{mic} = 0.66$ nm) for higher surface concentrations (¾Γ$_{mx}$, and ⅞Γ$_{mx}$). These values of $R_{mic}$ are physically unrealistic as they are less than half the length of an extended surfactant molecule. For this reason, model calculations similar to those described above were also made for surface aggregates of different geometries, viz., patch-like, ellipsoidal and wormlike micelles. The most acceptable morphology was oblate ellipsoids, with now two structural parameters, $R_n$ and $R_{lat}$, which also define the orientation of the micelle on the surface: the minor semi-axis $R_n$ is the axis in the direction perpendicular to the surface and defines the height of the micelle center above the silica surface. The major semi-axis $R_{lat}$ characterizes the lateral extension of the oblate surface micelles on the surface. It is assumed that the surface aggregates interact only through excluded volume interactions. The positioning of the micelles is then performed as with the spherical micelles, and again the Debye formula is employed to determine the micelle-micelle structure factor.

A subtlety of this procedure is that at first sight it seems to be incorrect, as it uses the formalism of separation into form factor and structure factor which is valid only for monodisperse objects of spherical symmetry. We show in the Appendix that it can be used in our case to calculate the scattered intensity, as before with resolution function.

The evidence for non-spherical surface micelles of $C_{12}DAO$ on silica nanoparticles suggested a comparison with the micelle shape in solution. A comparison of the scattering curves of $C_{12}DAO$ in $H_2O/D_2O$ in the absence and presence of silica II is shown in Figure 3. The data for the aqueous solution of $C_{12}DAO$ can be represented by a model of oblate ellipsoids, with $a = 1.47$ nm and $b = 2.47$ nm (where $a$ is the rotational semi-axis and $b$ the equatorial semi-axis of the ellipsoid core). The model of oblate micelles fits the data at high $q$ somewhat better than the prolate ellipsoidal model (fit not shown), as indicated by the lower residual, which was ~1.2 for the oblate ellipsoid, but ~1.6 for the prolate ellipsoid. Figure 3 also shows that the scattering curves of $C_{12}DAO$ in the absence and presence of the silica overlap in the high-$q$ region, indicating that the shape of the micelles in solution and at the surface is similar. Accordingly, the parameters $R_n$ and $R_{lat}$ of the surface micelle model were set to 1.5 nm and 2.2 nm, respectively, for all surface concentrations and the number of micelles, $N_{mic}$ was taken as the only adjustable parameter. Fits for $C_{12}DAO$ on silica II and silica III are shown in Figure 6. The good fit of the data in the entire $q$ range supports the chosen model of isolated oblate surface micelles. Furthermore, the values of $N_{mic}$ derived from the fits (Table 5) are similar to those found in the simple geometric analysis (Table 4).

To gain a better understanding of the way in which the calculated scattering function is influenced by the structural parameters $R_n$ and $R_{lat}$ we have varied them in a systematic manner, keeping one of them (and $N_{mic}$) fixed and varying the value of the other in a range from 0.5 to 3.5 nm, as shown in Figure 7 for the scattering profile of $\frac{7}{8}\Gamma_{mx}$ of $C_{12}DAO$ on silica II. Figure 7a shows the effect of a variation of $R_n$, at fixed $R_{lat} = 2.2$ nm and $N_{mic} = 94$. As can be seen, $R_n$ is directly related to the size of the ellipsoidal aggregates because decreasing its value to 0.5 nm or increasing its value to 3.5 nm causes a shift of $q_{max}$ to higher or lower values. $R_{lat}$ is related to the ordering of the micelles on the surface. This is indicated in Fig. 7b by the fact that a decrease of $R_{lat}$ from 2.2 to 0.5 nm causes a deformation of the size of the surface aggregates as their shell is no longer well defined. By increasing $R_{lat}$ to 3.5 nm a strong oscillation appears at $q \approx 1.3$ nm$^{-1}$, indicating intermicellar repulsion between the adsorbed surface aggregates (see also Figs. 6 and 7 in ref. 6).

## 4. Discussion

**4.1 Influence of surfactant head group**

Aggregate structures of surfactants in solution can often be predicted on the basis of the critical packing parameter $V/l_c a_0$, which expresses the preferred interface curvature of the aggregate in terms of molecular parameters alkyl chain volume $V$, the alkyl chain length $l_c$ and the head group area $a_0$.[21] $C_{12}$DAO in its cationic form at low pH is expected to have a higher effective head group area than in its nonionic form at high pH where the absence of electrostatic head group repulsion allows a closer packing of the head groups. Hence it should be possible to study the effect of head group size on the shape of surface aggregates simply by changing pH. However, this is not possible in the present case because pH has to be fixed near pH 9 to prevent flocculation of the silica dispersion. We have checked that added electrolyte has no effect on the morphology of the adsorbed layer of $C_{12}$DAO (Figure 3) under the experimental conditions, as expected for nonionic surfactants. It would be of interest to study the effect of electrolyte on the surfactant aggregates in the cationic form of the surfactant, but again this is not possible at the given pH of the system.

In our earlier study,[7] spherical surface aggregates were observed for the surfactant $C_{12}E_5$ on silica I, in line with the large head-group size of this molecule. Since $C_{12}$DAO has a much smaller head group than $C_{12}E_5$, one expects that surface aggregates of smaller curvature are preferred for this surfactant. This conjectured behavior is indeed found for $C_{12}$DAO on silica II and III, where we find oblate-shaped surface aggregates. On the other hand, spherical surface aggregates of $C_{12}$DAO are found on silica I, and they have similar dimensions as those of $C_{12}E_5$ on silica I. These findings suggest that the head group size (or packing parameter) of the surfactant is not the dominating factor for the *shape* of surface aggregates on the silica particles. However, the nature of the surfactant head group may have a pronounced influence on *the number of surface aggregates* per particle ($N_{mic}$). This is suggested by a comparison of the results for $C_{12}$DAO on silica I with the earlier results for $C_{12}E_5$ on the same silica.[7] For instance, at a surface concentrations ¾$\Gamma_{mx}$ we find $N_{mic} = 36$ for $C_{12}$DAO (Table 4), but $N_{mic}$ = 72 for $C_{12}E_5$.[7] The larger number for $C_{12}E_5$ may be attributed to its ability to form more than one strong hydrogen bond to surface silanol groups. However, other factors may also affect the number of surface aggregates, as is suggested by the fact that rather small numbers of surface aggregates ($N_{mic}$ up to 15) were found in the study of Triton X-100 (a technical-grade alkylphenyl polyoxyethylene surfactant) on a commercial silica sol (Bindzil B30). Since Triton X-100 is similar to $C_{12}E_5$ and the mean particle size of Bindzil B30 ($R_S$ = 7.7 nm) is similar to that of silica I, the large difference in $N_{mic}$ between these two systems is not clear.

**4.2 Influence of nanoparticle size**

An interesting finding of this study is the morphological transition from spherical to ellipsoidal shape of the surface aggregates. This transition must be caused by some property of the silica particles, either their surface chemistry or surface roughness, or the particle size. Since silica I and silica II were prepared by the same method, their surface properties are similar, as indicated by the similar zeta potentials (Table 1) and surface roughness ($a_s/a_{geom}$ in Table 3) of these two samples, when compared to silica III. Because the transition in surface aggregate shape occurs from silica I to silica II, we may conclude that it is not induced by changes in surface properties but by the increase in particle size. In the preceeding study[7] we conjectured that the formation of spherical surface aggregates of $C_{12}E_5$ on silica I was caused by the high surface curvature of the silica nanoparticles, which prevents an effective packing of the hydrophobic tails of the molecules in a bilayer configuration. This argument may be generalized by noting that the formation of surface aggregates at concentrations below the *CMC* depends on favorable interactions of the surfactant heads with the solid surface, and that the morphology of the surface aggregates will be determined by a balance of amphiphile-amphiphile (A-A) and amphiphile-surface (A-S) interactions. At weakly convex surfaces (large silica particles), micellar aggregates of relatively low mean curvature can have favourable A-S interactions without significant changes in aggregate structure, i.e. without sacrificing A-A interaction energy. This appears to be the situation for the oblate-shaped surface micelles of $C_{12}$DAO on silica II and silica III. On the other hand, at highly convex surfaces (small silica particles), optimization of the A-A and A-S interactions may favor smaller surface aggregates, if that leads to a larger number of surface contacts per unit area of the solid particle – even at the cost of a higher A-A bending energy. This seems to the situation for $C_{12}$DAO on silica I. The transition from spherical to oblate shape of the surface aggregates can then be seen as a relaxation from a strained to an unstrained curvature of the surface aggregates, since oblate micelles represent the favored aggregate shape of $C_{12}DAO$ in the bulk solution. The present work suggests that this morphological transition occurs at a particle radius $R_S \approx 10\text{-}12$ nm, i.e. $R_{mic}/R_S \approx 0.2$. To our knowledge, no theoretical model for this morphological transition exists in the literature, but such a model would be most valuable for gaining a better understanding of this phenomenon.

**Conclusion**

SANS has been used to study the shape of surface aggregates of $C_{12}DAO$ formed at the surface of spherical silica nanoparticles with diameters from 16 to 42 nm. In agreement with results for other nonionic surfactants studied earlier (Triton X-100[5] and $C_{12}E_5$[7]) it is found that $C_{12}DAO$ does not form a laterally uniform adsorbed layer on the surface of the silica nanoparticles, but rather they form small surface aggregates. The present work presents evidence for a morphological transition of the surface micelles as a function of the particle size of the silica nanoparticles. Spherical surface aggregates are formed on particles of 16 nm, but oblate ellipsoidal surface micelles on silica particles of 27 and 42 nm diameter. The formation of spherical surface micelles of $C_{12}DAO$ on the surface of the smallest silica particles is favored because this kind of morphology optimizes the surface free energy by increasing the contact area between the surface micelles and the silica surface. This energy decreases as the surface curvature of the silica particles decreases with increasing particle size, and thus aggregates of lower curvature, such as oblate ellipsoids are favored on larger silica particles. These ellipsoidal surface aggregates are characterized by a minor semi-axis $R_n$ and a major semi-axis $R_{lat}$, which also defines the effective surface area of the surface aggregates. The dimensions of the ellipsoidal surface aggregates are similar to those of $C_{12}DAO$ micelles in the aqueous solutions. From a comparison of the present results with those of the preceding studies it is concluded that the shape of the surface aggregates (spherical or ellipsoidal) is not determined by the size of surfactants head group. However, for spherical surface micelles it appears that the nature of the head group can have a strong influence on the maximum number of surface aggregates per particle. Specifically, the maximal number of surface micelles of $C_{12}DAO$ is much smaller than for $C_{12}E_5$ at the same silica, presumably due to its less hydrophilic character. However, further systematic work is needed to clarify the role of surfactant head group – surface interactions on the type of surface aggregates and the maximum number density of aggregates on the silica particles. We are planning such studies in our laboratories.

The present work shows that the micelle-decorated silica model provides a reliable and versatile basis for determining size and shape and the number of surface aggregates of amphiphiles on spherical nanoparticles from SANS scattering data. The results obtained on the basis of this form-factor model are consistent with those derived by a simple geometric analysis of the Guinier and Porod regimes of the SANS data.

**Acknowledgments**

The authors wish to thank D. Berger for help with the Transmission Electron Microscopy. D.L. is grateful to Deutscher Akademischer Austauschdienst (DAAD) and the Fundación Gran Mariscal de Ayacucho (Fundayacucho) for receiving a doctoral scholarship, and to TU Berlin for a Promotionsabschluß-Stipendium. Financial support by DFG through project FI 235/16-2 and the cooperation initiated in the framework of the French-German Network "Complex Fluids: From 3 to 2 Dimensions" (Project FI 235/14-3) is also gratefully acknowledged.

# Appendix

**Scattering from index-matched silica particles decorated with ellipsoidal micelles**

Micelles adsorbed on the surface of a sphere are confined to a spherical shell. Correlations between their centers of mass are generated by excluded volume and possibly other interactions between the micelles. In our experiments, silica nanoparticles are index-matched, i.e., they do not contribute to the signal. Scattering from interacting spherical micelles can be factorized in a product of form and structure factor:

$$I(q) = \frac{N}{V} S(q) P(q) \tag{A1}$$

where $P(q)$ is the form factor related to the scattering amplitude $F(q)$, which is the Fourier transform of the scattering length density distribution, by

$$P(q) = \left\langle |F(q)|^2 \right\rangle = |\langle F(q) \rangle|^2 \tag{A2}$$

The last equality is a direct consequence of spherical symmetry, the average being (also) a rotational one. The case of interacting ellipsoidal micelles can be deduced in the same way as equation (A1) by simply keeping $\langle |F(q)|^2 \rangle$ and $|\langle F(q) \rangle|^2$ separate:

$$I(q) = \frac{N}{V}\left[S(q)|\langle F(q)\rangle|^2 + \langle |F(q)|^2\rangle - |\langle F(q)\rangle|^2\right] \qquad (A3)$$

In order to see if the simpler factorization can be used, it is thus necessary to calculate $\langle |F(q)|^2 \rangle$ and $|\langle F(q) \rangle|^2$. In the Figure A.1, the result for an isolated oblate ellipsoid at 1 vol.-%, with a contrast of $\Delta\rho = 5.1 \; 10^{10} \text{cm}^{-2}$, is shown.

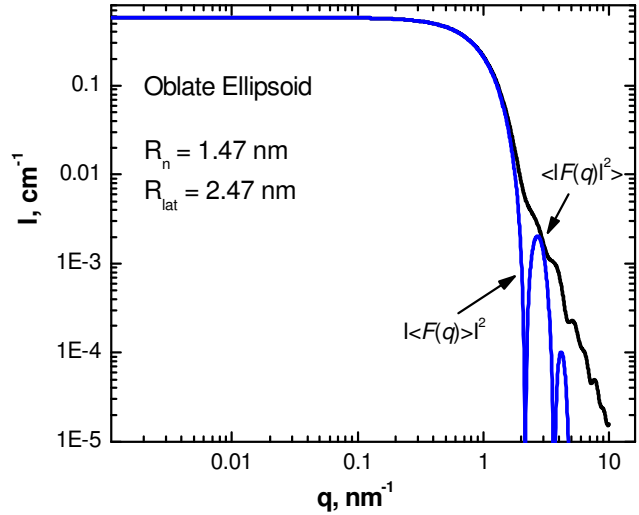

Fig. A.1. Comparison of $\langle |F(q)|^2 \rangle$ and $|\langle F(q) \rangle|^2$ for isolated oblate ellipsoids.

As can be seen in Figure A.1, $\langle |F(q)|^2 \rangle$ and $|\langle F(q) \rangle|^2$ coincide for small angles ($q < 1.5 \text{ nm}^{-1}$), but differences appear at $q > 1.5 \text{ nm}^{-1}$ due to stronger oscillations of the square of the average of the amplitude $F(q)$. One thus has to discuss the low-$q$ and high-$q$ regions separately:

 - At low $q$, the two functions are the same, and equation (A3) reduces to (A1).

 - At high $q$, the structure factor tends to one, and equation (A3) reduces to the measured form factor scattering $P(q) = \langle |F(q)|^2 \rangle$.

Given that the maxima of $S(q)$ in our experimental case are located well below 1 nm$^{-1}$, and Porod surface (*i.e.* form factor) scattering is observed above 1.5 nm$^{-1}$, it is clear that $S(q) = 1$ in this high-$q$ range. Thus, equation (A1) can be safely used with the experimentally measured form factor for ellipsoidal micelles of typical size 1.5 to 2.5 nm, adsorbed and interacting on the surface of an indexed-matched silica sphere of much larger radius. Limits of this approach may be a weaker separation in length scales, *e.g*. caused by a very high surface density of ellipsoidal micelles, which is not the case here. Note that we have applied such calculations to the similar case of interacting cylindrical (albeit non adsorbed) micelles.[22]

# Tables

**Table 1.** Characterization of the silica nanoparticles*

| Silica Sol | $R_{TEM}$ nm | $SD_{TEM}$ | $\zeta$ mV | $\kappa$ µm cm$^{-1}$ |
|---|---|---|---|---|
| I | 8.3 | 0.126 | -45.6 | 121 |
| II | 13.7 | 0.104 | -43.2 | 66 |
| III | 21.3 | 0.027 | -32.7 | 69 |

*average particle radius $R$ and size distribution $SD$ from TEM, zeta potential $\zeta$ and conductivity $\kappa$ of the silica dispersion

**Table 2.** Characterization of the silica dispersions by SANS*:

| Silica Sol | $R_S$ (nm) | $\sigma$ | $\langle R_S \rangle$ (nm) | $\langle A_S \rangle$ (nm$^2$) | $\langle V_S \rangle$ (nm$^3$) |
|---|---|---|---|---|---|
| I | 8.20 | 0.10 | 8.24 | $8.62 \cdot 10^2$ | $2.42 \cdot 10^3$ |
| II | 13.50 | 0.10 | 13.57 | $2.34 \cdot 10^3$ | $1.08 \cdot 10^4$ |
| III | 21.00 | 0.10 | 21.11 | $5.65 \cdot 10^3$ | $4.06 \cdot 10^4$ |

*Parameters of the log-normal size distribution, $R_S$ and $\sigma$, average radius $\langle R_S \rangle$, average surface area $\langle A_S \rangle$, average volume $\langle V_S \rangle$ of the silica beads.

**Table 3.** Surface characterization of the silicas.*

| Silica Sol | $a_s$, m$^2$ g$^{-1}$ | $a_{geom}$, m$^2$ g$^{-1}$ | $a_s/a_{geom}$ |
|---|---|---|---|
| I | 190 | 166 | 1.14 |
| II | 134 | 101 | 1.33 |
| III | 120 | 65 | 1.85 |

*Specific surface area $a_s$ from the BET analysis; geometrical surface area $a_{geom}$ from the particle radius and mass density of silica.

**Table 4.** Characteristics of the surfactant layer adsorbed at the silica nanoparticles derived from the SANS data and from the adsorption isotherm*

| Sample | SANS | | | Adsorption Isotherm | |
|---|---|---|---|---|---|
| | $V_{dry}$ ($10^3$ $nm^3$) | $\delta$(nm) | $N_{mic}$ | $V_{dry}$ ($10^3$ $nm^3$) | $\delta$(nm) |
| **Silica I** | | | | | |
| ¼$\Gamma_{mx}$ | 0.16 | 0.18 | 4 | 0.42 | 0.47 |
| ½$\Gamma_{mx}$ | 0.70 | 0.76 | 17 | 0.84 | 0.90 |
| ¾$\Gamma_{mx}$ | 1.51 | 1.50 | 36 | 1.26 | 1.28 |
| ⅞$\Gamma_{mx}$ | 1.74 | 1.69 | 42 | 1.47 | 1.47 |
| **Silica II** | | | | | |
| ¼$\Gamma_{mx}$ | 0.91 | 0.39 | 22 | 1.14 | 0.48 |
| ½$\Gamma_{mx}$ | 2.29 | 0.93 | 55 | 2.28 | 0.93 |
| ¾$\Gamma_{mx}$ | 3.28 | 1.34 | 81 | 3.43 | 1.36 |
| ⅞$\Gamma_{mx}$ | 3.98 | 1.55 | 101 | 4.00 | 1.56 |
| **Silica III** | | | | | |
| ⅞$\Gamma_{mx}$ | 6.20 | 1.06 | 150 | 9.67 | 1.62 |

*dry volume $V_{dry}$ and effective thickness $\delta$ of the adsorbed $C_{12}$DAO layer; $N_{mic}$ is the number of surface micelles

**Table 5.** Parameters of the micelle-decorated silica model for $C_{12}$DAO on silica particles*

| Sample | $R_{mic}$ (nm) | $N_{mic}$ |
|---|---|---|
| **Silica I (spherical)** | | |
| ½$\Gamma_{mx}$ | 1.63 | 38 |
| ¾$\Gamma_{mx}$ | 1.97 | 36 |
| ⅞$\Gamma_{mx}$ | 1.97 | 42 |
| **Silica II (oblate)** | | |
| ¼$\Gamma_{mx}$ | 1.5 / 2.2 | 26 |
| ½$\Gamma_{mx}$ | 1.5 / 2.2 | 55 |
| ¾$\Gamma_{mx}$ | 1.5 / 2.2 | 87 |
| ⅞$\Gamma_{mx}$ | 1.5 / 2.2 | 94 |
| **Silica III (oblate)** | | |
| ⅞$\Gamma_{mx}$ | 1.5 / 2.2 | 155 |

* Silica I: Best-fit values of $R_{mic}$ and $N_{mic}$ for spherical surface micelles. Silica II and silica III: best-fit values of $N_{mic}$ for oblate ellipsoidal surface micelles of fixed values of $R_n$ and $R_{lat}$
.

# Figures

**Figure 1.** TEM images and particles size distribution histograms for silica I, silica II and silica III. The histogram are based on the diameters of at least of 200 different particles from different TEM images.

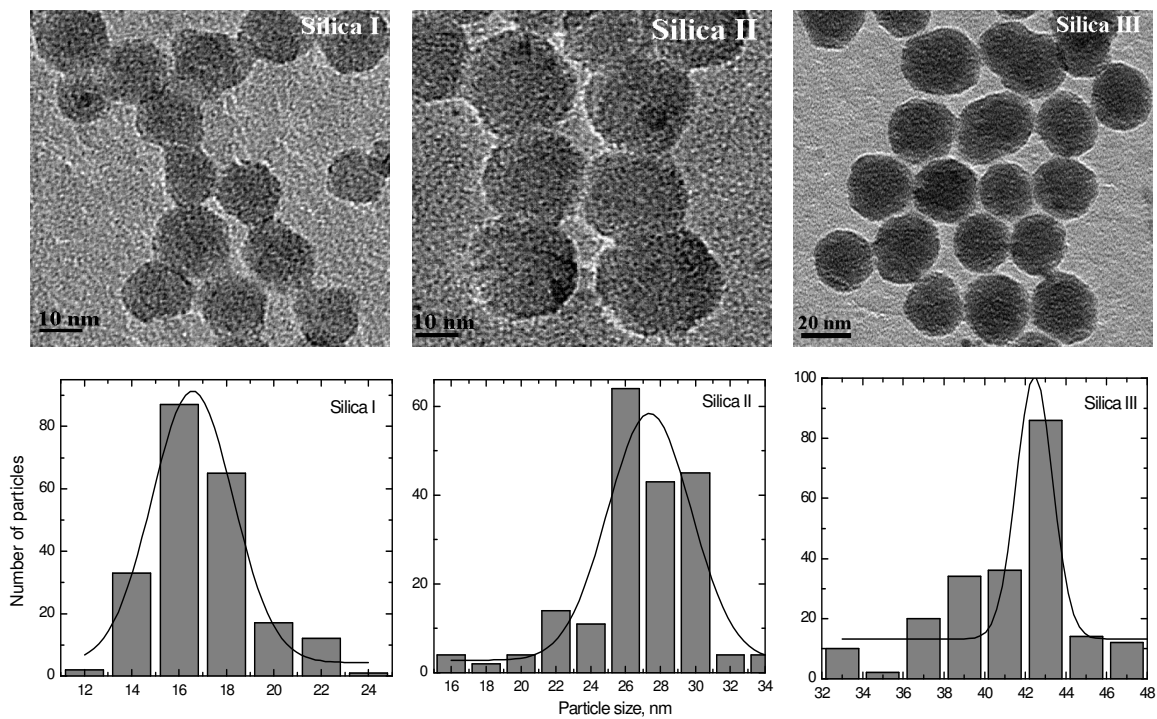

**Figure 2**. Scattering profile $I(q)$ of a dilute dispersion of silica II in nearly pure $D_2O$ at pH 9 (298 K). The solid line represents a fit with the log-normal size distribution function. The inset shows $I(q)$ for this silica in contrast matching $H_2O/D_2O$ to indicate the quality of the contrast match. The constant background arises mostly from incoherent scattering from protons of $H_2O$.

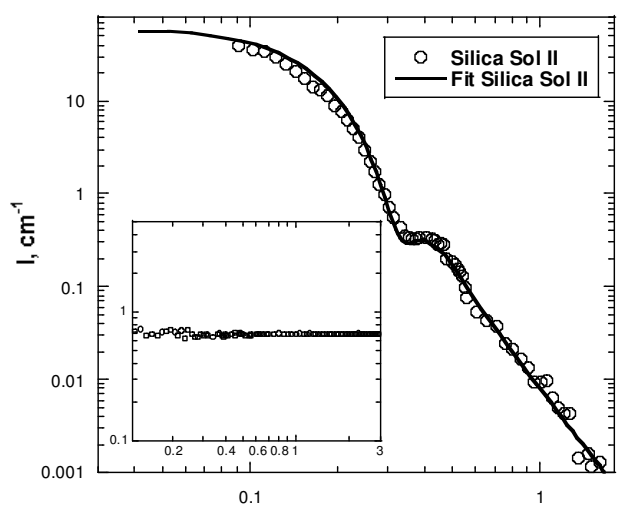

**Figure 3.** SANS profiles $I(q)$ for $C_{12}DAO$ in the contrast matching $H_2O/D_2O$ in the absence and presence of silica II. Also shown is the scattering profile for the silica containing system with added 0.1 M NaBr. The scattered intensities were normalized with volume fraction $\varphi$ of the surfactant in the system.

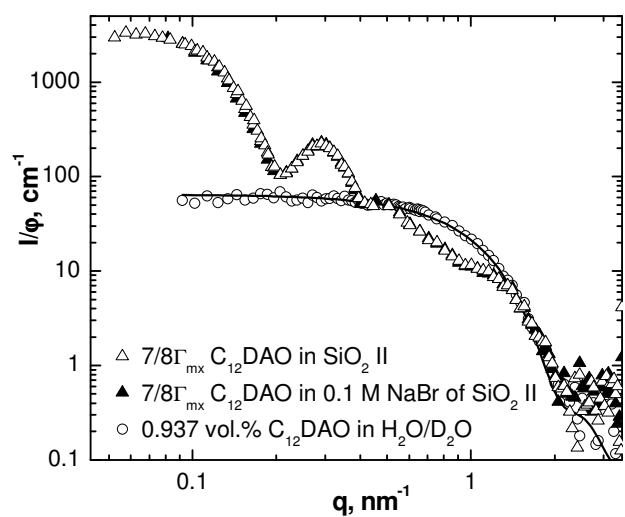

**Figure 4.** Experimental SANS profiles $I(q)$ and intensities predicted by the spherical core-shell model with shell thicknesses $\delta = 1.6$, 3.2 and 4.0 nm for surface concentration $\tfrac{7}{8}\Gamma_{mx}$ of $C_{12}DAO$ on silica II in contrast matching $H_2O/D_2O$. The inset shows the predicted intensities with thickness 3.2 and 4.0 nm multiplied with scale factors $f = 0.22$ and 0.12, respectively.

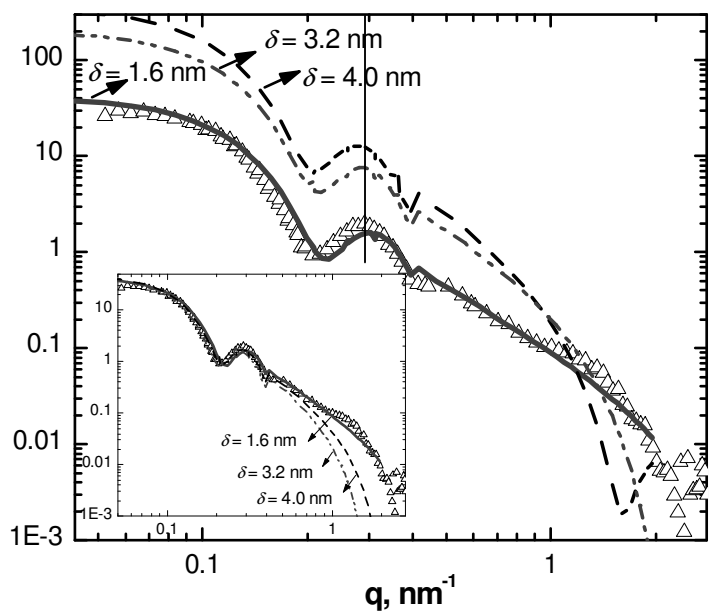

**Figure 5.** SANS profiles $I(q)$ for silica I with adsorbed $C_{12}DAO$ (surface concentrations $\tfrac{7}{8}\Gamma_{mx}$ and $\tfrac{3}{4}\Gamma_{mx}$) and fits by the micelle-decorated silica model (solid curves) for spherical surface micelles of radius 2 nm (parameters see Table 5). Results for surface concentration $\tfrac{7}{8}\Gamma_{mx}$ are shifted relative to that for $\tfrac{3}{4}\Gamma_{mx}$ by factor of 4.

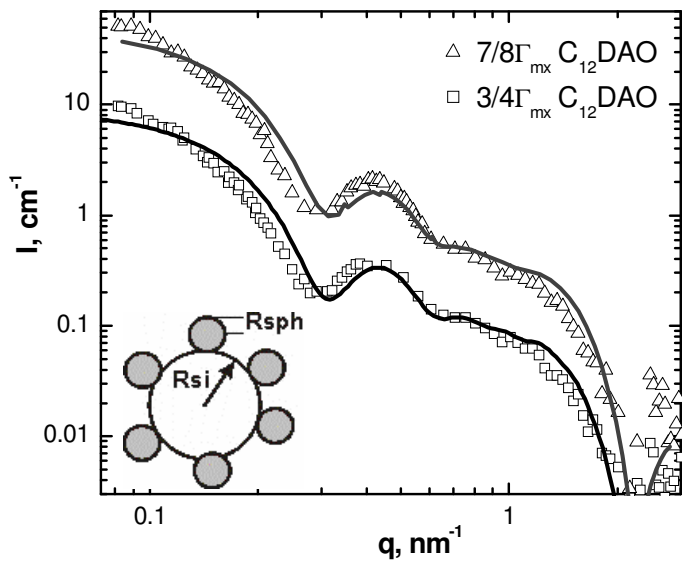

**Figure 6.** SANS profiles $I(q)$ for silica II and silica III with adsorbed $C_{12}DAO$ in contrast-matching $H_2O/D_2O$ and fits by the micelle-decorated silica model for ellipsoidal micelles (solid curves): (a) silica II at surface concentrations ¼$\Gamma_{mx}$, ½$\Gamma_{mx}$, ¾$\Gamma_{mx}$ and ⅞$\Gamma_{mx}$; (b) silica III with surface concentrations ⅞$\Gamma_{mx}$ of $C_{12}DAO$ (parameters see Table 5). In (a) the curves for higher surface concentrations are shifted vertically relative to that of ¼$\Gamma$mx by factors of 2 (½$\Gamma_{mx}$), 4 (¾$\Gamma_{mx}$) and 14 (⅞$\Gamma_{mx}$).

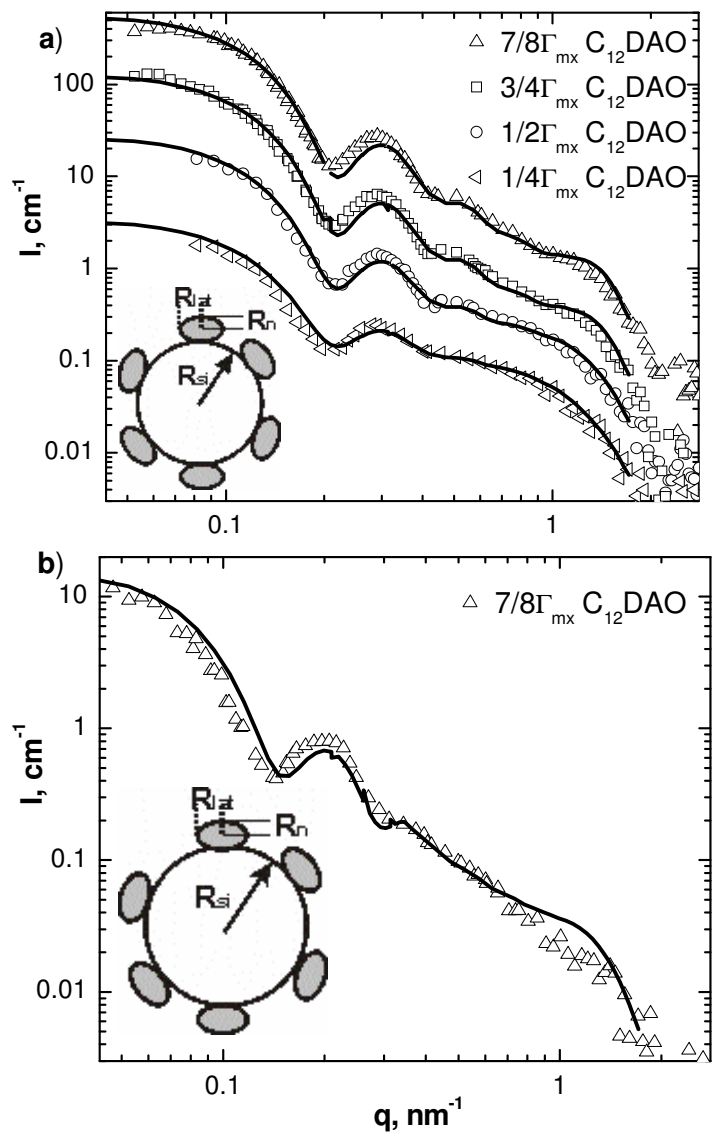

**Figure 7.** Scattering profile for $C_{12}DAO$ (surface concentration $\frac{7}{8}\Gamma_{mx}$) on silica II and results predicted by the micelle-decorated silica model for ellipsoidal surface micelles: (a) varying the normal semi-axis $R_n$, at fixed $R_{lat}$ = 2.2 nm and $N_{mic}$ = 94; (b) varying the lateral semi-axis $R_{lat}$, at fixed $R_n$ =1.5 nm $N_{mic}$ = 94.

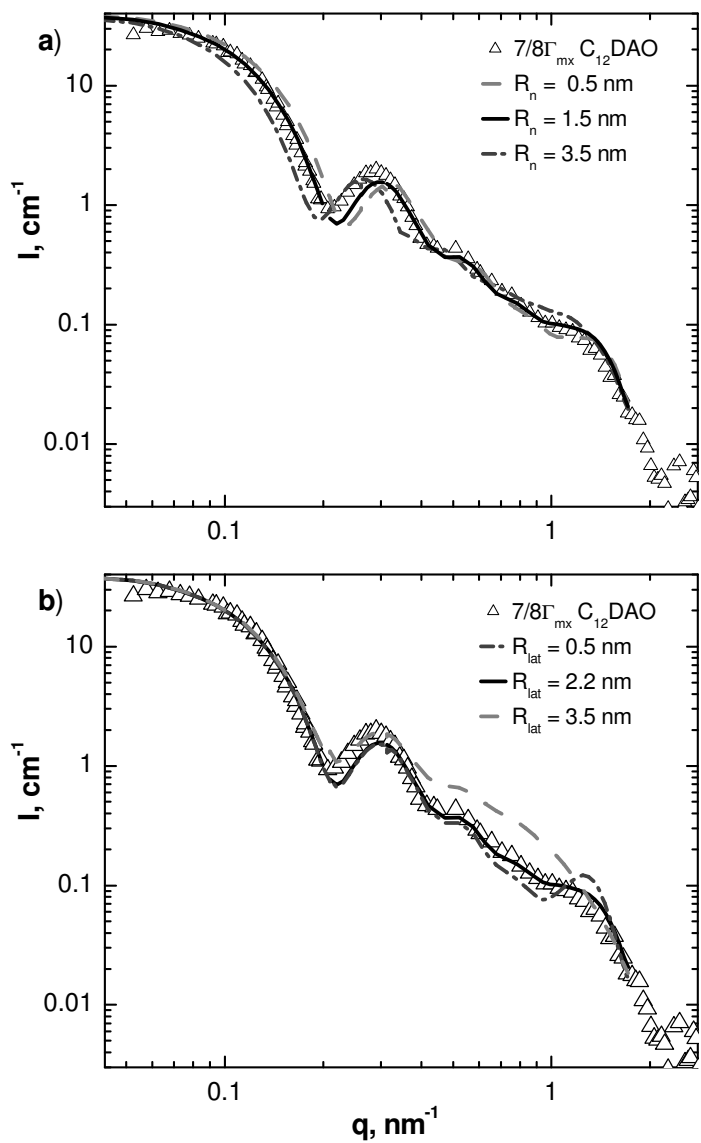

**References**

1 Dietsch, O.; Eltekov, A.; Bock, H.; Gubbins, K.E.; Findenegg, G.H. *J. Phys. Chem. C* **2007**, *111*, 16045-16054.
2 (a) Grant, L.M.; Tiberg, F.; Ducker, W.A. *J. Phys. Chem. B* **1998**, *102*, 4288–4294. (b) Grant, L. M.; Ederth, T.; Tiberg, F. *Langmuir* **2000**, *16*, 2285-2291. (c) Blom, A.; Duval, F. P.; Kovács, L.; Warr, G. G.; Almgren, M.; Kadi, M.; Zana, R. *Langmuir* **2004**, *20*, 1291–1297.
3 (a) Cummins, P.G.; Staples, E.; Penfold, J. *J. Phys. Chem.* **1990**, *94*, 3740-3745. (b) Cummins, P.G.; Staples, E.; Penfold, J. *J. Phys. Chem.* **1991**, *95*, 5902-5905. (c) Cummins, P.G.; Penfold, J.; Staples, E. *J. Phys. Chem.* **1992**, *96*, 8092–8094.
4 Penfold, J.; Staples, E.; Tucker, I.; Cummins, P.G. *J. Phys. Chem.* **1996**, *100*, 18133–18137.
5 Despert, G.; Oberdisse, J. *Langmuir* **2003**, *19*, 7604–7610.
6 Oberdisse, J. *Phys. Chem. Chem. Phys.* **2004**, *6*, 1557-1561.
7 Lugo, D.; Oberdisse, J.; Karg, M.; Schweins, R.; Findenegg, G.H. *Soft Matter* **2009**, *5*, 2928–2936.
8 Li, X.; Lin, Z.; Cai, J.; Scriven; L.A. Davis, H,T *J. Phys. Chem.* **1995**, *99*, 10865-10878.
9 Menge, U.; Lang, P.; Findenegg, G.H.; Strunz, P. *J. Phys. Chem. B* **2003**, *107*, 1316-1320.
10 Sterpone, F.; Marchetti, G.; Pierleoni, C.; Marchi, M. *J. Phys. Chem. B* **2006**, *110*, 11504-11510.
11 (a) Chernik, G. G.; Sokolova, E. P. *J. Colloid Interface Sci.* **1991**, *141*, 409-414. (b) Benjamin, L. *J. Phys. Chem.* **1964**, *68*, 3575-3581. (c) Kresheck, G. C. *J. Am. Chem. Soc.* **1998**, *120*, 10964-10969. (d) Timmins, P. A.; Hauk, J.; Wacker, T.; Welte, W. *Febs Letters* **1991**, *280*, 115-120. (e) Timmins, P. A.; Leonhard, M.; Weltzien, H. U.; Wacker, T.; Welte, W. *Febs Letters* **1988**, *238*, 361-368. (f) Barlow, D. J.; Lawrence, M. J.; Zuberi, T.; Zuberi, S. *Langmuir* **2000**, *16*, 10398-10403.
12 Király, Z.; Findenegg, G. H. *Langmuir* **2000**, *16*, 8842-8849.
13 (a) Pettersson, A.; Rosenholm, J. B. *Langmuir* **2002**, *18*, 8436-8446. (b) Pettersson, A.; Rosenholm, J. B. *Langmuir* **2002**, *18*, 8447-8454.
14 Stöber, W.; Fink, A.; Bohn, E. *J. Colloid Interface Sci.* **1968**, *26*, 62-69.
15 Brûlet, A.; Lairez, D.; Lapp, A.; Cotton, J.-P. *J. Appl. Cryst.* **2007**, *40*, 165-xxx; Lindner, P. *in Neutrons, X-rays and Light: Scattering Methods Applied to Soft Condensed Matter* **2002**, *ed. P. Lindner and Th. Zemb, Boston, 1st edn., ch. 2, pp 23-48.*
16 Oberdisse, J.; Deme, B. *Macromolecules* **2002**, *35*, 4397-4405.
17 Lowell, S.; Shields, J.E.; Thomas, M.A.; Thommes, M. *Characterization of Porous Solids and Powders: Surface Area, Pore Size and Density*, **2004**, Kluwer Academic Publishers, Dordrecht
18 Tanford, C. *J. Phys. Chem.* **1972**, *76*, 3020-3024.
19 Holmberg, K.; Jönsson, B.; Kronberg, B.; Lindman, B. *Surfactants and Polymers in Aqueous Solution* **2003**, *2nd ed.*, John Wiley & Sons.
20 Pusey, P.N. in *Neutrons, X-rays and Light: Scattering Methods applied to Soft Condensed Matter*, ed. B.J. Gabrys, **2000**, *1st ed.,* Gordon and Breach Science Publishers, Netherlands, chap. 4, pp. 77-102.
21 Israelachvili, J.N.; Mitchell, D.J.; Ninham, B.W. *J. Chem. Soc. Faraday Trans. 2* **1976**, *72*, 1525-1568.
22 Oberdisse, J.; Regev, O.; Porte, G. *Journal of Physical Chemistry B* **1998**, *102*, 1102-1108.